\documentclass[intlimits,twoside,a4paper]{article}

\usepackage{amsmath,amssymb}
\usepackage{graphicx}

\usepackage[T2A]{fontenc}
\usepackage[cp1251]{inputenc}

\usepackage[eqsecnum]{cmpj2}

\issue{2014}{17}{4}{43004}
\doinumber{10.5488/CMP.17.43004}

\title[Gas--liquid phase equilibrium in ionic fluids]%
{Gas--liquid phase equilibrium in ionic fluids:  Coulomb  versus non-Coulomb interactions}
\author{O. Patsahan}

\address{Institute for Condensed Matter Physics of the National
Academy of Sciences of Ukraine, \\ 1 Svientsitskii St., 79011 Lviv,
Ukraine}

\date{Received August 28, 2014, in final form November 10, 2014}
\authorcopyright{O. Patsahan, 2014}

\begin{document}

\maketitle

\begin{abstract}
Using the collective variables theory, we study the effect of competition between Coulomb and dispersion forces on the gas--liquid phase behaviour of a model ionic fluid, i.e. a charge-asymmetric primitive  model with additional short-range attractive interactions.  Both the critical parameters and the coexistence envelope are calculated   in a one-loop approximation as a function of the parameter $\alpha$  measuring the relative strength of the Coulomb to short-range interactions.
We found   the very narrow region of $\alpha$ bounded from the both sides by  tricritical points which separates the models with   ``nonionic'' 
and ``Coulombic'' phase behaviour.
This is at variance with the result of available computer simulations where no tricritical point is found for the finely-discretized lattice version of the model.

\pacs 05.70.Fh, 64.60.De, 64.60.Kw
\end{abstract}

\section{Introduction}
Critical and phase behaviour of ionic fluids has been intensively studied for the last decades. These studies were
stimulated by  earlier  experiments on ionic solutions that yielded
three types of  critical behavior:   mean-field and   Ising-like
behavior as well as   crossover between the two
\cite{singh,levelt1}. In accordance with these peculiarities, ionic solutions were
conventionally divided into two classes, namely: ``solvophobic''
systems with Ising-like critical behavior in which  Coulomb forces are not supposed to play a major
role (the solvent is generally characterized by high dielectric
constant) and
``Coulombic'' systems in which the phase separation is primarily
driven by Coulomb interactions (the solvent is characterized by
low dielectric constant) \cite{levelt1,Weingartner-Merkel}.  The critical behaviour  of the Coulombic systems has been a subject of great debates.  At present, the analysis of the existing
experimental data for Coulombic and solvophobic systems has shown that the asymptotic critical behaviour is Ising-like for both classes
\cite{Gutkowskii-Anisimov,Schroer:12}.
Experiments that supported the expectation of
mean-field critical behaviour
could not be reproduced in later works \cite{wiegand-98,wiegand-98-2,Schroer:06}. Strong
evidence for the Ising universal class for the Coulombic fluids has been also found by recent
simulations \cite{caillol-levesque-weis:02,Hynnien-Panagiotopoulos,luijten,kim-fisher-panagiotopoulos:05}
and theoretical studies \cite{patsahan:04:1,ciach:06:1,Parola-Reatto:11}.  More recent
accurate experiments indicate  in general a crossover to the
mean-field behaviour,
where the non-classical region increases with the polarity of the solvent
\cite{Schroer:review,Sengers_Shanks:09,Schroer:12}. Theoretical predictions of such behaviour has been  recently given
in \cite{Patsahan:13, Patsahan:14}.

In this paper we continue our systematic study of the phase behaviour of ionic fluids. Our aim  is to study the
effect of the strength of electrostatic interactions (the dielectric constant of the solvent) on the phase separation in
ionic solutions.
It should be noted that ionic and nonionic (``classical'') fluids  differ greatly  in  the strength of the interactions. In simple neutral
fluids, the interaction energy  is of the order of the thermal energy, i.e., the reduced
temperature $T^{*}=k_\textrm{B}T/\phi_\textrm{min}\simeq 1$ ($\phi_\textrm{min}$ is the depth of the interaction potential at its minimum).
By contrast, in typical molten salts,  the interionic interactions are more than one order of magnitude
larger than $k_\textrm{B}T$, i.e., $T^{*}\ll 1$.  In electrolyte solutions, the strength  of the Coulomb interactions depends on the
dielectric constant of the solvent. In addition, the phase diagrams of Coulomb dominated fluids are quite asymmetric when compared
to nonionic fluids \cite{Weingartner-Kleemeier,Weiss_Schroer:08,Schroer_Vale:09}.

The evolution of the gas--liquid phase diagram of the charged Yukawa fluid with an increase  of
ionic interactions  was studied  in      \cite{Kristof_Boda:03}
 using both the mean-spherical approximation and the Gibbse ensemble Monte Carlo simulations.
It has been found
that  the gas--liquid coexistence envelope  changes from a ``nonionic type''
(such as that of the one-component  Yukawa fluid) to a ``Coulombic type''(such as that of the primitive model) when the strength of
the Coulomb interactions is increased.
However, the issue
of  the watershed between the nonionic and ionic phase behaviour (the ``solvophobic''  and ``Coulombic'' systems) has not
been addressed so far.
Here, we address  this  issue using a simple model of ionic fluids, i.e., the charge-asymmetric primitive  model (PM) supplemented by
short-range attractive interactions where the short-range attraction
is considered as an approximation to van der Waals interactions
\cite{Patsahan:13,Kristof_Boda:03,ciach_stell:2001,Ciach_Stell:2002,ciach:05,Diehl_Panagiotopoulos:03}.
The model without the Coulomb interactions (a hard-sphere square-well model) exhibits a  gas--liquid coexistence
typical of nonionic (``solvophobic'') fluids.
Another limiting model, i.e., the PM, demonstrates the  ``Coulombic'' phase diagram.
A theoretical background for this
study is the statistical field theory that exploits the method of collective variables (CVs) \cite{zubar,jukh,Yuk-Hol,Pat-Mryg-CM}.
The theory enables us to
derive an exact expression for the functional of grand partition
function  of the model and on this basis to develop the
perturbation theory
\cite{Pat-Mryg-CM,patsahan-mryglod-patsahan:06}.
The
well-known approximations for the free energy, in particular
Debye-H\"{u}ckel limiting law and the mean spherical approximation,
can be reproduced within the framework of this theory \cite{patsahan-mryglod-patsahan:06}. Links between
this approach and the  field theoretical approach
\cite{Ciach-Gozdz-Stell-07} were established in
\cite{patsahan-mryglod:06} for the case of the RPM.
Using the CVs  theory we have recently derived a microscopic-based  effective Hamiltonian for the  model with short- and
long-range interactions \cite{Patsahan:13}. A distinguishing feature of the developed approach is that it is enables one to
obtain   the  coefficients of the Hamiltonian, including the square-gradient term,
within  the framework of the same approximation. For the above-mentioned model, we have found explicit expressions for all
the relevant coefficients in the one-loop approximation \cite{Patsahan:13}.
 For the free energy, this approximation
coincides with the well-known random-phase approximation (RPA) and, thus, produces the mean-field phase diagram. Here,
we use the effective  Hamiltonian derived in  \cite{Patsahan:13} to study the gas--liquid phase behaviour of
the model with parameters ranging from the purely   non-Coulombic regime to the purely  Coulombic regime.

The paper is arranged as follows. A theoretical background is given in section~2. The results for the gas--liquid critical parameters and
phase diagrams are presented in section~3.  Concluding remarks are made in section~4.

\section{Theory}

\subsection{Model}
We start with  a   two-component model of ionic fluids.
The pair interaction potential is assumed to be of the following form:
\begin{equation}
U_{\alpha\beta}(r)=\phi_{\alpha\beta}^\textrm{HS}(r)+\phi_{\alpha\beta}^\textrm{SR}(r)+\phi_{\alpha\beta}^\textrm{C}(r),
\label{2_1}
\end{equation}
where  $\phi_{\alpha\beta}^\textrm{HS}(r)$ is the interaction potential
between the two additive hard spheres of diameters $\sigma_{\alpha}$ and $\sigma_{\beta}$. The potential $\phi_{\alpha\beta}^\textrm{SR}(r)$
 describes the short-range  (van der Waals-like) attraction.
$\phi_{\alpha\beta}^\textrm{C}(r)$ is the
Coulomb potential:
$\phi_{\alpha\beta}^\textrm{C}(r)=q_{\alpha}q_{\beta}\phi^\textrm{C}(r)$, where
$\phi^\textrm{C}(r)=1/(\epsilon r)$ and $\epsilon$ is the dielectric constant of the solvent.
The ions of the species $\alpha=1$ carry  an electrostatic charge
$q_{+}=q_{0}$ and those of species $\alpha=2$ carry an opposite charge $q_{-}=-zq_{0}$ ($q_{0}$ is an
elementary charge and $z$ is the parameter of charge asymmetry).  Overall charge neutrality requires that
$q_{+}N_{+}+q_{-}N_{-}=0$, where $N_{+}$ and $N_{-}$ are, respectively,
the number of positive and negative ions. In general, the binary system of hard spheres  interacting via the potential
$\phi_{\alpha\beta}^\textrm{SR}(r)$  can exhibit  both the gas--liquid  and demixion phase transitions.  Thus, we simplify the model assuming
that  (i) the hard spheres are of the same diameter  $\sigma_{\alpha}=\sigma_{\beta}=\sigma$ and (ii)
$\phi_{11}^\textrm{SR}(r)=\phi_{22}^\textrm{SR}(r)=\phi_{12}^\textrm{SR}(r)=\phi^\textrm{SR}(r)$. With these restrictions,  the uncharged system can only exhibit
a gas--liquid phase separation
and a possible demixion is ruled out. Then,  we specify   $\phi^\textrm{SR}(r)$ in the form of  the square-well (SW) potential of
depth $\varepsilon$  and range $\lambda$. The system of hard spheres interacting
through the SW potential with $\lambda=1.5\sigma$ can serve as a reasonable model for simple (nonionic)
fluids.

It is worth noting that in the
treatments of models with hard cores, the perturbation potential is
not defined uniquely inside the hard core. Here, we  use the
Weeks-Chandler-Andersen   regularization scheme  for the  both potentials $\phi^\textrm{C}(r)$ and
$\phi^\textrm{SR}(r)$  \cite{wcha}. In this case,  the Fourier transforms of these potentials have the form:
\begin{equation}
\widetilde \phi^\textrm{SR}(k)=-\varepsilon\sigma^3 \frac{4\pi}{x^3}[-\lambda x
~\cos(\lambda x) + \sin(\lambda x)]
\label{phi-sr-k}
\end{equation}
and
\begin{equation}
\widetilde\phi^\textrm{C}(k)=4\pi\sigma^3\displaystyle\frac{\sin(x)}{\epsilon
x^{3}}\,,
\label{wca_coul}
\end{equation}
where
$x=k\sigma$.

\subsection{Effective Hamiltonian near the gas--liquid critical point}

We consider  the model (\ref{2_1})--(\ref{wca_coul}) near the gas--liquid critical point.
Using the method of collective variables, we can present the effective Ginzburg-Landau Hamiltonian of the model
as follows (see \cite{Patsahan:13} and references therein):
\begin{eqnarray}
{\cal H}^\textrm{eff}&=&a_{1,0}\rho_{0,N}+\frac{1}{2!\langle
N\rangle}\sum_{{\mathbf{k}}}\left(a_{2,0}+k^{2}a_{2,2}\right)\rho_{{\bf
k},N}\rho_{-{\bf k},N}+\frac{1}{3!\langle
N\rangle^{2}}\sum_{{\mathbf{k}}_{1},{\mathbf{k}}_{2}} a_{3,0}
\nonumber \\
&& \times\rho_{{\bf k_{1}},N}\rho_{{\bf k_{2}},N}\rho_{-{\bf
k_{1}}-{\bf k_{2}},N}+\frac{1}{4!\langle
N\rangle^{3}}\sum_{{\mathbf{k}}_{1},{\mathbf{k}}_{2},{\mathbf{k}}_{3}}a_{4,0}\rho_{{\bf
k_{1}},N}\rho_{{\bf k_{2}},N}\rho_{{\bf k_{3}},N}\rho_{-{\bf
k_{1}}-{\bf k_{2}}-{\bf k_{3}},N}+\ldots,
\label{H_eff}
\end{eqnarray}
where $\rho_{{\mathbf k},N}$ is the collective variable (CV) which describes the value of the $\mathbf
k$-th fluctuation mode of the total number density. In a one-loop approximation, the coefficients of Hamiltonian (\ref{H_eff}) have the  form:
\begin{eqnarray}
&&a_{1,0}=-\Delta\nu_{N}+\beta\rho\widetilde\phi^\textrm{SR}(0)-\widetilde{C}_{1,\text{C}}\,,
\label{a10_rpa}\\
&& a_{2,0}=-\rho\,\widetilde {C}_{2,\textrm{HS}}+\beta\rho\widetilde\phi^\textrm{SR}(0)-\rho\,\widetilde
{C}_{2,\text{C}}\,,
\label{a20_rpa}
\\
&& a_{2,2}=-\frac{1}{2}\rho\,\widetilde {C}_{2,\textrm{HS}}^{(2)}+\frac{1}{2}\beta\rho\widetilde\phi^{\textrm{SR},(2)}
-\frac{1}{4 \langle N\rangle}\sum_{\mathbf{q}}\widetilde
g^{(2)}(q)\left[1+\widetilde g(q)\right]\,,
\label{a22_rpa}
\\
&&a_{n,0}=-\rho^{n-1}\,\widetilde {C}_{n,\textrm{HS}}-\rho^{n-1}\,\widetilde {C}_{n,\text{C}}, \qquad
n\geqslant 3\,.
\label{an0_rpa}
\end{eqnarray}
Here, we introduce the following notations. The subscript $\textrm{HS}$ refers to the hard-sphere system and  the superscript $(2)$ in
equation (\ref{a22_rpa}) denotes the second-order derivative with the respect of the wave vector $k$:
$\widetilde
g^{(2)}(q)=\partial^{2}\widetilde
g(|\mathbf{q}+\mathbf{k}|)/
\partial k^{2}\rvert_{k=0}$,
$\widetilde {C}_{2,\textrm{HS}}^{(2)}=\partial^{2} \widetilde
{C}_{2,\textrm{HS}}(k)/\partial k^{2}\rvert_{k=0}$, and
$\widetilde\phi^{\textrm{SR},(2)}=\partial^{2} \widetilde
\phi^\textrm{SR}(k)/\partial k^{2}\rvert_{k=0}$.

The addend $\Delta\nu_{N}$ in equation~(\ref{a10_rpa})  is related to the chemical potentials
\begin{equation}
\Delta\nu_{N}=\nu_{N}-\nu_{N,\textrm{HS}}\,,
\qquad
\nu_{N}=\frac{z\bar\nu_{1}+\bar\nu_{2}}{1+z},
\label{delta_nu}
\end{equation}
where $\bar\nu_{\alpha}$ is determined by
\begin{equation}
\bar \nu_{\alpha}=\nu_{\alpha}+\frac{\beta}{2V}\sum_{{\mathbf
q}}\left[\widetilde\phi^\textrm{SR}(q)+q_{\alpha}^{2}\widetilde\phi^\textrm{C}(q)\right],
\label{nu_alpha}
\end{equation}
$\nu_{\alpha}$ is the dimensionless chemical potential,
$\nu_{\alpha}=\beta\mu_{\alpha}-3\ln\Lambda_{\alpha}$ and
$\mu_{\alpha}$ is the chemical potential of the $\alpha$th species;
$\beta=1/k_\textrm{B}T$ is the reciprocal temperature; $\Lambda_{\alpha}^{-1}=(2\pi m_{\alpha}\beta^{-1}/h^{2})^{1/2}$ is
the inverse de Broglie thermal wavelength.

$\widetilde{C}_{n,\textrm{HS}}$ is the Fourier
transform of the $n$-particle direct correlation function of a
one-component hard-sphere system at $k=0$, $\rho=\langle N\rangle/V=\rho_{1}+\rho_{2_{}}$ is the total number density.
Explicit expressions for   $\widetilde{C}_{n,\textrm{HS}}$ and
$\widetilde {C}_{2,\textrm{HS}}^{(2)}$ for $n\leqslant 4$ in the
Percus-Yevick (PY) approximation are given in
reference \cite{Patsahan:13}  (see appendix in \cite{Patsahan:13}).

The last term on the right-hand side of equations~(\ref{a10_rpa})--(\ref{an0_rpa}) results from  the  charge-charge
correlations being taken into account through  integration over the charge subsystem  \cite{Patsahan:13}.
In particular, $\rho^{n-1}\widetilde{C}_{n,\text{C}}$ reads
\begin{eqnarray}
\rho^{n-1}\widetilde{C}_{n,
\text{C}}&=&\frac{(n-1)!}{2}\frac{1}{\langle
N\rangle}\sum_{\mathbf{q}}\left[\widetilde g(q)\right]^{n},
\label{Cn_C}
\end{eqnarray}
where
\begin{eqnarray}
\widetilde g(q)&=&-\frac{\beta\rho \widetilde\phi^\textrm{C}(q)}{1+\beta\rho
\widetilde\phi^\textrm{C}(q)}\,.
\label{g_q}
\end{eqnarray}
 All the coefficients in equation (\ref{H_eff}) except $a_{2,2}$ describing the square-gradient
term can be obtained from the one-loop (RPA) free energy \cite{Patsahan:13}
\begin{eqnarray*}
 a_{n,0}=\rho^{n-1}\frac{\partial^{n}(-\beta f_{\textrm{RPA}})}{\partial
 \rho^{n}}=\rho^{n-1}\widetilde{C}_{n}(0, \ldots), \qquad
 n\geqslant 2,
\end{eqnarray*}
where $\widetilde{C}_{n}(0, \ldots)$ denotes the Fourier
transform of the n-particle  direct correlation function of the full
system in the long-wavelength limit. It is worth noting that the
functions $\widetilde{C}_{n}$ differ from the ordinary direct
correlation functions $\widetilde c_{n}$ by an ideal term
\cite{hansen_mcdonald}. Equation $a_{2,0}=0$ yields a spinodal curve. In combination  with equation $a_{3,0}=0$ it determines the critical point.

The square-gradient term $a_{2,2}$ given by equation (\ref{a22_rpa})   is also derived  in the one-loop approximation
(see \cite{Patsahan:13} for  details). It is essential that the last addend on the right-hand side of (\ref{a22_rpa})
describes the  short range attraction which arises from the integration over the charge subsystem. We  emphasize  that
although the original Hamiltonian of the purely Coulombic model [$\phi_{\alpha\beta}^\textrm{SR}(r)=0$] does not include direct
pair attractive interactions
of number density fluctuations, the effective short-range attraction does appear in the effective Hamiltonian  (\ref{H_eff}).
It is worth noting that  the expression for $a_{2,2}$ given by
(\ref{a22_rpa})  produces a correct result for the density correlation length $\xi$ in the limit  of charged point particles
(see \cite{Patsahan:13} and references herein).

Coefficient $ a_{1,0}$ is the excess part of the chemical potential
$\nu_{N}$ connected with the short-range attractive and long-range
Coulomb interactions. Equation $a_{1,0}=0$ yields  the expression for the chemical potential in
the RPA.

It is worth noting that there is no difference between the
charge-asymmetric PM and the restricted primitive model (RPM) having $z=1$ at this level of approximation
\cite{stell1,levin-fisher,Cai-Mol1,patsahan-mryglod-patsahan:06}.
Hereafter, we briefly refer to the model (\ref{2_1})--(\ref{wca_coul})  as a RPM-SW model.

\section{Gas--liquid phase diagram}

Here, we study the gas--liquid phase diagram of the the RPM-SW model for the model parameters ranging from the RPM limit to the SW limit. Using equations (\ref{phi-sr-k})--(\ref{wca_coul}) and
(\ref{delta_nu})--(\ref{g_q}) we rewrite the expressions for the coefficients of the effective Hamiltonian [see (\ref{a10_rpa})--(\ref{an0_rpa})]
as follows:
\begin{eqnarray}
a_{1,0}&=& -\nu_{N}+\nu_{N,\textrm{HS}}+\frac{1}{2\alpha T^{\text{C}}}-\frac{1}{2T^{\text{C}}} -\frac{8\eta\lambda^{3}}{\alpha T^{\text{C}}}+i_{1}\,, \label{a1_sw+pm} \\
a_{2,0}&=&-\rho\,\widetilde {C}_{2,\textrm{HS}}-\frac{8\eta\lambda^{3}}{\alpha T^{\text{C}}}+i_{2}\,, \label{a20_sw+pm}\\
a_{2,2}&=&-\frac{1}{2}\rho\,\widetilde
{C}_{2,\textrm{HS}}^{(2)}+\frac{4}{5}\frac{\eta\lambda^{5}}{\alpha T^{\text{C}}}-\frac{1}{6\pi T^{\text{C}}}i_{12}\,, \label{a22_sw+pm}\\
a_{n,0}&=&-\rho^{n-1}\widetilde {C}_{n,\textrm{HS}}+i_{n}\,, \qquad
n\geqslant 3\,, \label{an_sw+pm}
\end{eqnarray}
where \cite{Patsahan:13}
\begin{eqnarray}
i_{n}&=&\frac{(n-1)!(-\kappa^{2})^{n-1}}{\pi T^{\text{C}}}
\int_{0}^{\infty} x^{2}\left[\frac{\sin(x)}{x^{3}+\kappa^{2}\sin(x)}\right]^{n}{\rm
 d}x,
 \label{i_n}
 \\
i_{12}&=&\int_{0}^{\infty} x^{6}\Big\{\kappa^{2}x^{2}\left[1+\cos^{2}(x)\right] +2\left[x^{3}-2\kappa^{2}\sin(x)\right]\left[2x\cos(x)-3\sin(x)\right] \nonumber \\
&&
 +x^{5}\sin(x)\Big\}\big/\left[x^{3}+\kappa^{2}\sin(x)\right]^{4}{\rm
 d}x,
 \label{i_12}
\end{eqnarray}
and $\kappa=\kappa_\textrm{D}\sigma$  with $\kappa_\textrm{D}$ being the Debye number. In equations
(\ref{a1_sw+pm})--(\ref{i_12}),
 $T^{\text{C}}$ is the reduced temperature defined  as the ratio between the thermal energy $k_{\text{B}}T$ and the Coulomb energy
of the opposite charged hard spheres at  contact, $E^{\text{C}}=zq_{0}^{2}/\epsilon\sigma$,
\begin{equation}
 T^{\text{C}}=\frac{k_{\text{B}}T}{E^{\text{C}}}
 =\frac{k_{\text{B}}T\epsilon\sigma}{zq_{0}^{2}}\,,
\label{T-C}
\end{equation}
$\alpha$  is the ratio of the  Coulomb and square-well energies at contact
\begin{equation}
\alpha=\frac{E^{\text{C}}}{\varepsilon}=\frac{zq_{0}^{2}}{\epsilon\sigma\varepsilon}\,,
\label{alpha}
\end{equation}
and $\eta=\pi\rho\sigma^{3}/6$ is the packing fraction of the ions. For $\varepsilon$ being fixed, $\alpha$ measures the  strength of the Coulomb interactions. In this case,
either the increase of ion charges or  the decrease of  dielectric constant for fixed charges
results in the increase of the parameter $\alpha$. It is worth noting that $a_{n,0}$ with $n\geqslant 3$ is independent of $\alpha$
when the reduced temperature is given by equation~(\ref{T-C}). The approximation (\ref{a1_sw+pm})--(\ref{i_12}) produces
the mean-field phase diagram.
 \begin{figure}[!b]
 \centerline{
 \includegraphics[width=0.52\textwidth]{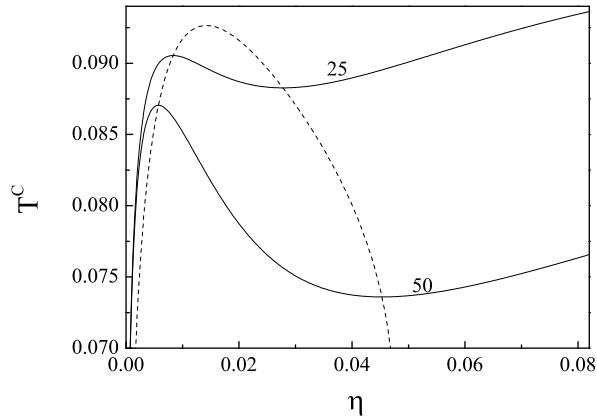}    
 }
 \caption{RPM-SW model: the loci of equations $a_{2,0}=0$
 (solid lines) and $a_{3,0}=0$ (dashed line) for $\alpha=25$ and $50$.
 $T^\textrm{C}$ is given by equation (\ref{T-C}),
 $\eta=\pi\rho\sigma^{3}/6$ is the packing fraction and $\alpha=q_{0}^{2}z/(\epsilon\sigma\varepsilon)$.} \label{fig1}
 \end{figure}

\begin{figure}[!t]
 \centerline{
 \includegraphics[width=0.47\textwidth]{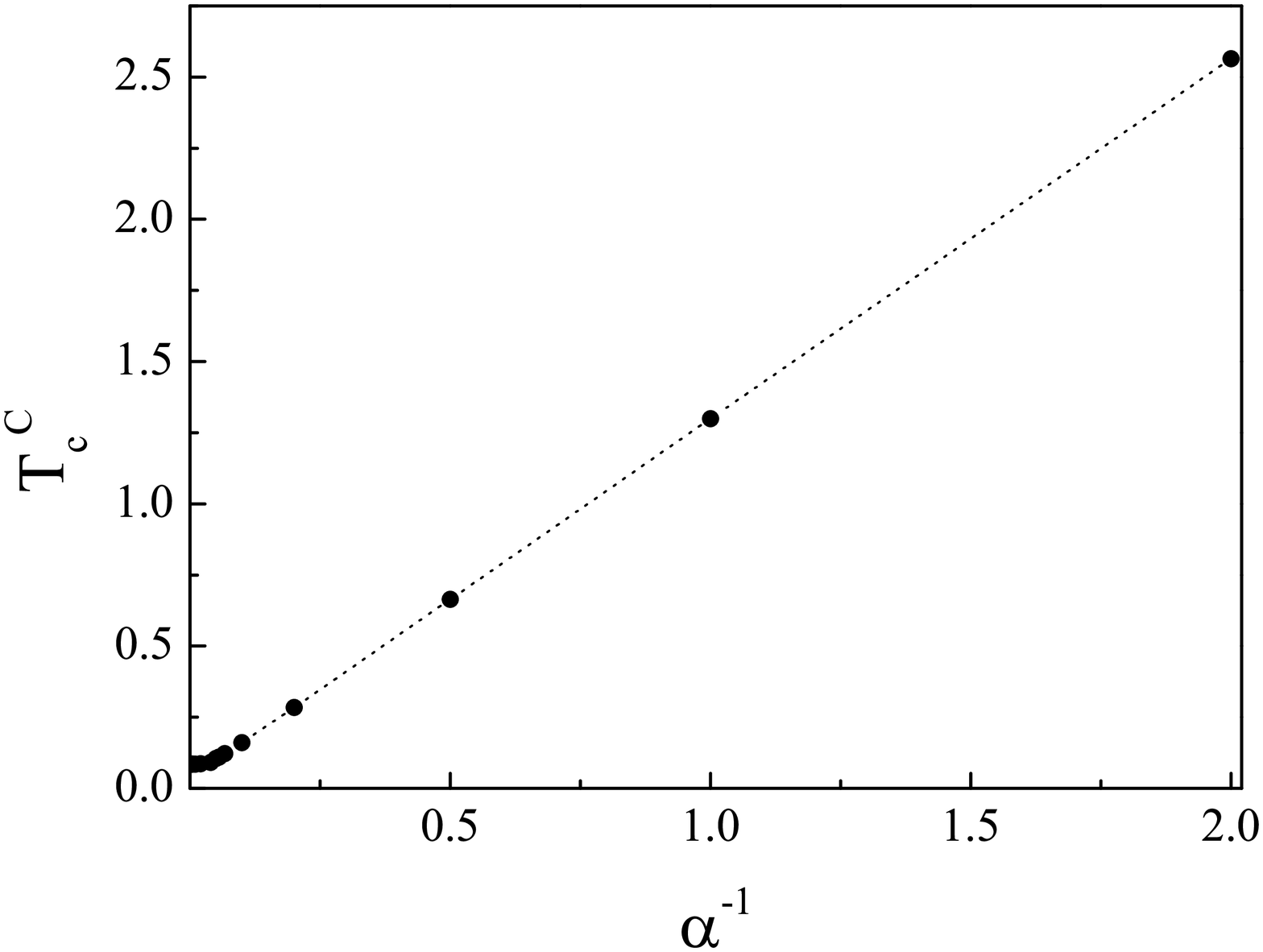}
\hfill
 \includegraphics[width=0.46\textwidth]{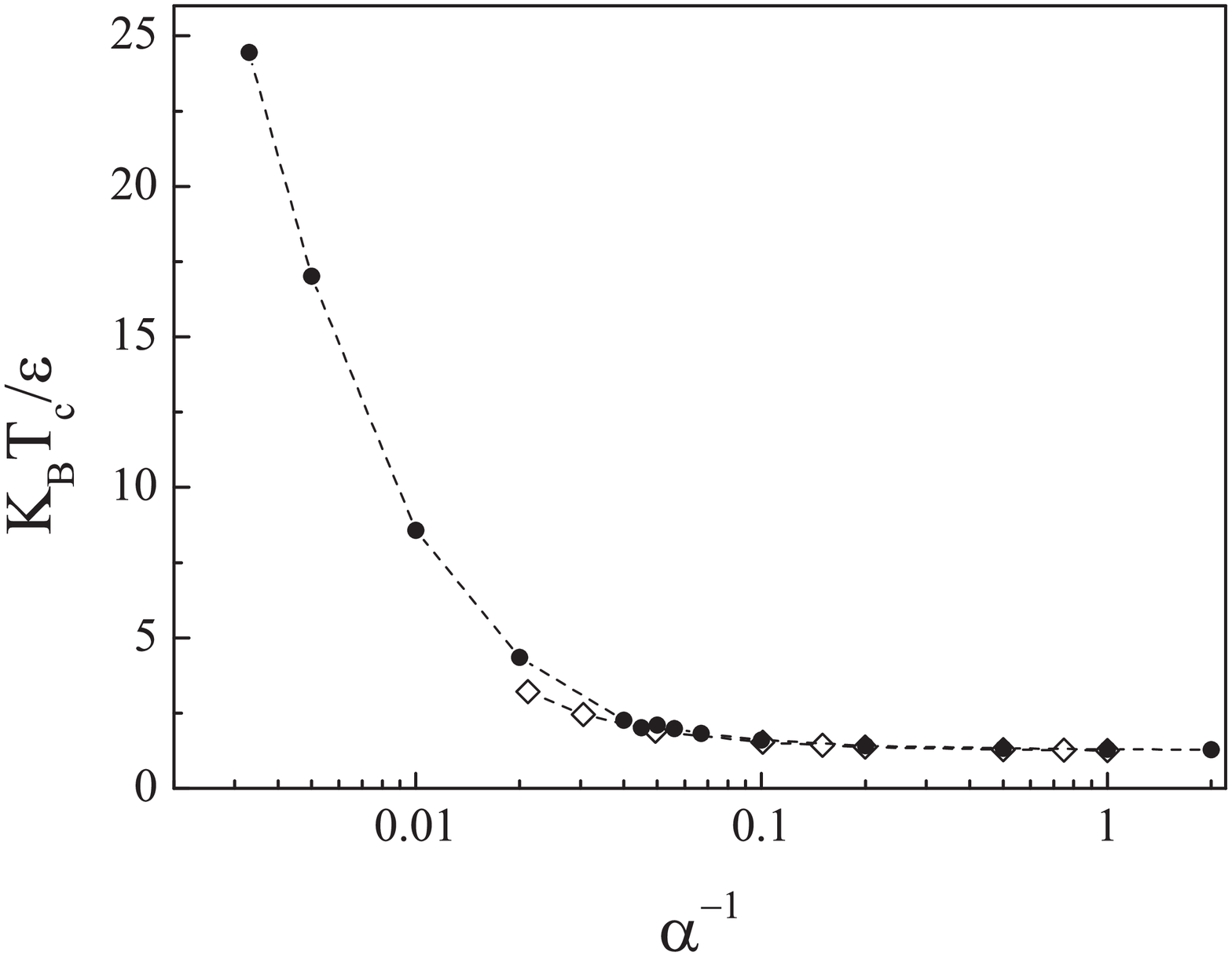}
}
\vspace{-3mm}
\parbox[t]{0.5\textwidth}{
 \caption{Reduced critical temperature $T_\textrm{c}^\textrm{C}$  of the RPM-SW model as a function of $\alpha^{-1}\sim\epsilon$
 [see equation (\ref{alpha})].
  The line is a guide to the eye.} \label{fig2}
  }
\parbox[t]{0.5\textwidth}{
 \caption{Reduced critical temperature $k_{\text{B}}T_\textrm{c}/\varepsilon$  of the RPM-SW model as a function of
 $\alpha^{-1}\sim\epsilon$
 [see equation (\ref{alpha})]. Full circles  are the results of the present work; open diamonds
 correspond  to the results  of simulations \cite{Diehl_Panagiotopoulos:03}.
The line is a guide to the eye.} \label{fig3}
  }
 \end{figure}

\begin{figure}[!b]
 \centerline{
 \includegraphics[width=0.47\textwidth]{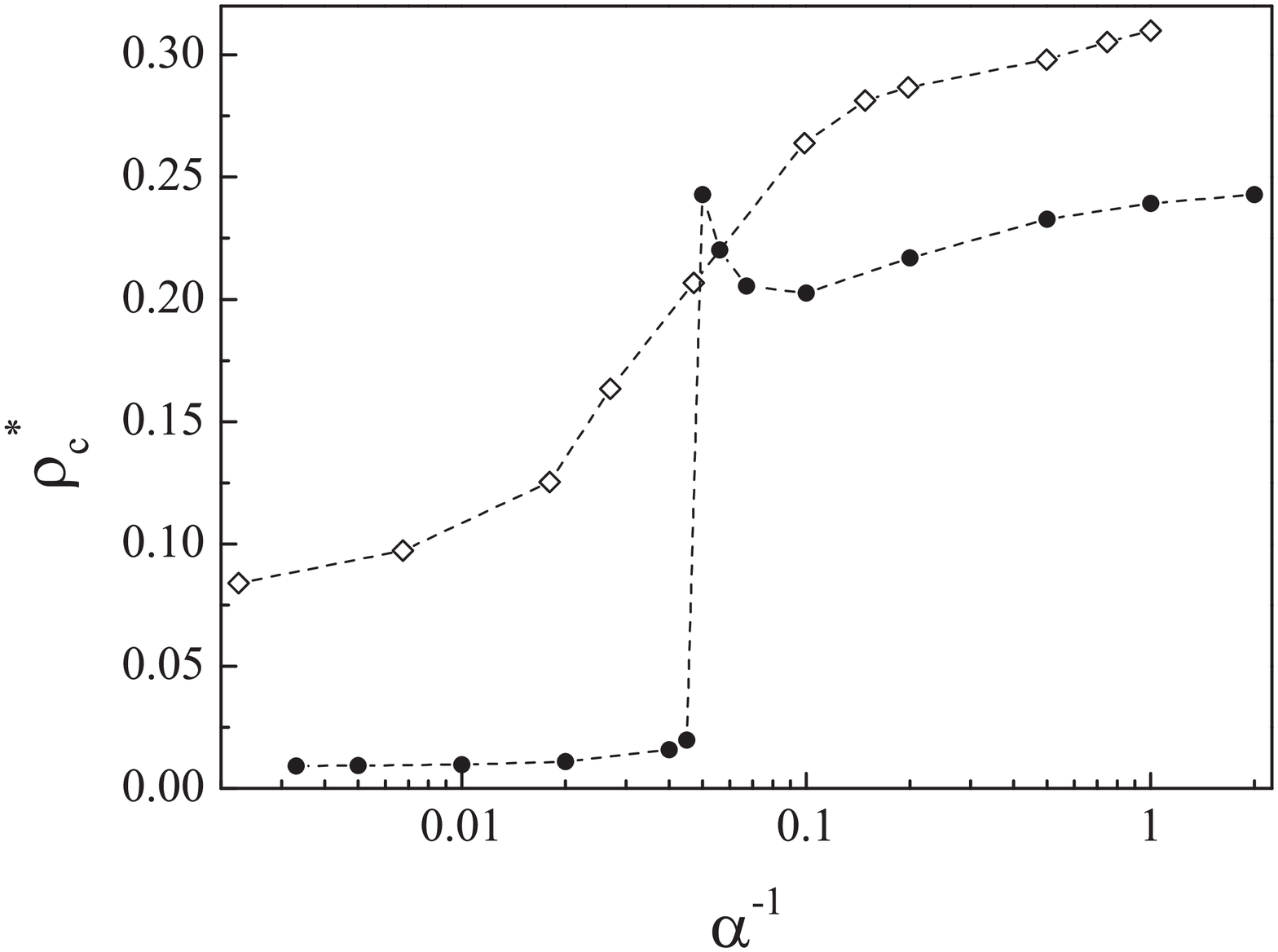}
 \hfill
 \includegraphics[width=0.46\textwidth]{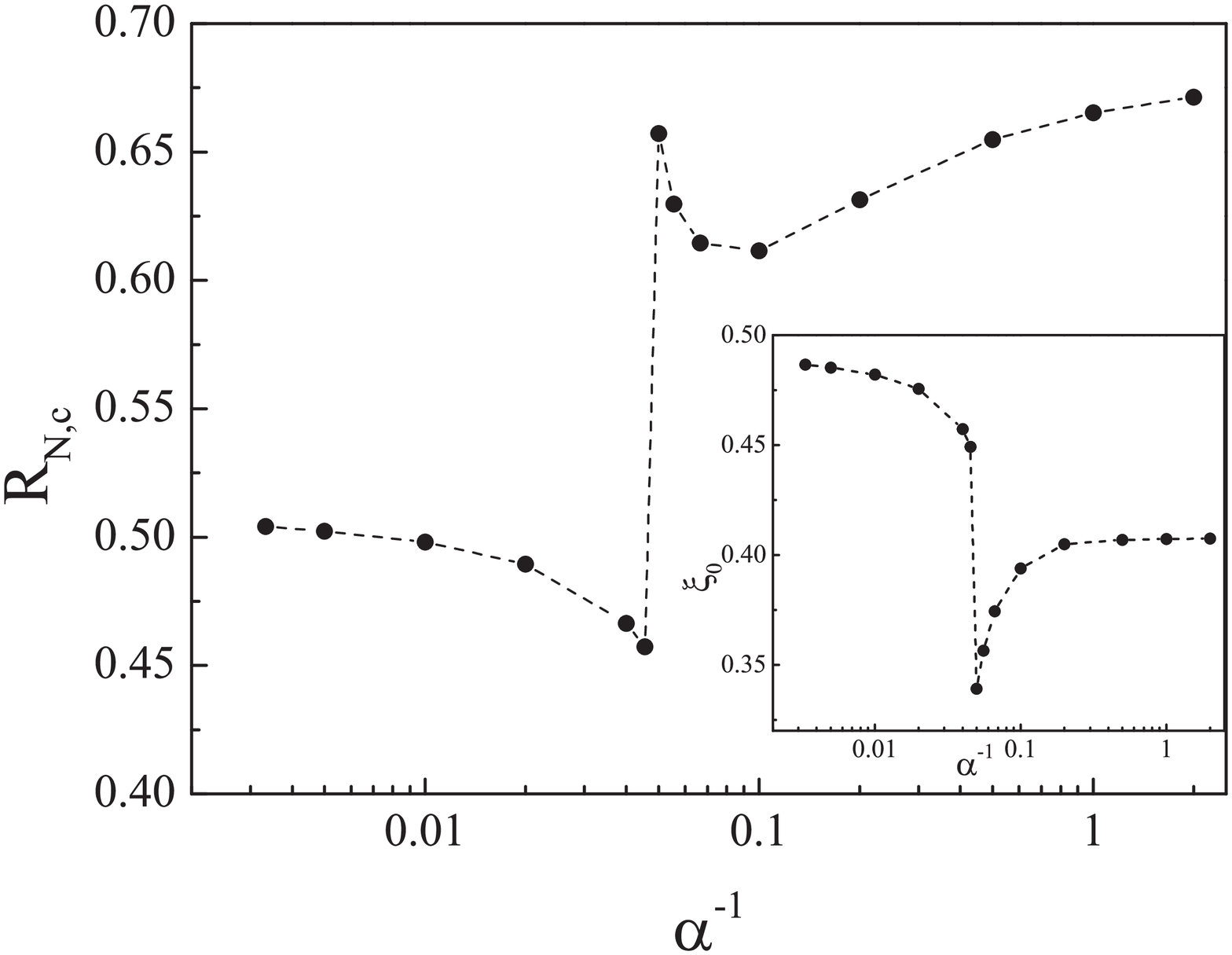}
}
\vspace{-3mm}
\parbox[t]{0.5\textwidth}{
 \caption{Reduced critical density $\rho^{*}_\textrm{c}=\rho_\textrm{c}\sigma^{3}$ of the RPM-SW model as
 a function of $\alpha^{-1}\sim\epsilon$
 [see equation (\ref{alpha})].  Full circles  are the results of the present work; open diamonds
 correspond  to the results  of simulations \cite{Diehl_Panagiotopoulos:03}. The line is a guide to the eye.
 } \label{fig4}
 }
 \parbox[t]{0.5\textwidth}{
 \caption{Dimensionless range of the effective density-density  interaction  $R_{N,\textrm{c}}=R_{N}(T_\textrm{c},\rho_\textrm{c})$
 of the RPM-SW model as
 a function of $\alpha^{-1}\sim\epsilon$
 [see equation (\ref{alpha})].  The inset shows the dimensionless  amplitude of the correlation length $\xi_{0}^{*}$ as a function of $\alpha^{-1}$. The line is a guide to the eye.
 } \label{fig5}
 }
 \end{figure}

First, we consider the critical point. At the critical point, the  system of equations
 \begin{eqnarray}
  a_{2,0}(\rho_\textrm{c},T_\textrm{c})=0, \qquad a_{3,0}(\rho_\textrm{c},T_\textrm{c})  =0
\label{cr-point}
 \end{eqnarray}
holds  yielding the critical temperature and the critical density for the fixed $\alpha$.  Generally, the curve given by the second equation of (\ref{cr-point})  intersects the spinodal twice, i.e., at its maximum and minimum points
(see figure~\ref{fig1}). The gas--liquid critical point is located at the maximum point of the spinodal.
The results for the gas--liquid critical parameters are displayed in figures~\ref{fig2}--\ref{fig4}.

In figure~\ref{fig2}, the dependence of the reduced critical temperature  given by (\ref{T-C}) on the parameter $\alpha^{-1}\sim\epsilon$ is shown.
As is seen, $T_\textrm{c}^{\text{C}}$ increases almost linearly with an increase of the dielectric constant.   The trend of $T_\textrm{c}^{\text{C}}$
agrees with the experimental observations when one expresses the critical temperature of real ionic solutions in the RPM values
\cite{Schroer:review,Wagner:04}. When the strength of the Coulomb interactions increases, ($\alpha^{-1}\rightarrow 0$) $T_\textrm{c}^{\text{C}}$ converges to the RPM limit $T_\textrm{c}^\textrm{C}=0.08446$ \cite{Patsahan:13}.

In order to compare our results with the available results of computer simulations \cite{Diehl_Panagiotopoulos:03} we rewrite
the reduced temperature in equation (\ref{T-C}) as $k_{\text{B}}T/\varepsilon$, the usual definition for the SW potential.
Figure~\ref{fig3} shows that the critical temperature scaled by the SW depth $\varepsilon$ rapidly decreases with a decrease of the Coulomb
interactions and then slowly approaches the SW limit $k_{\text{B}}T_\textrm{c}/\varepsilon=1.2667$ \cite{Patsahan:13}.
Such behaviour is in  agreement with the results of simulations  \cite{Diehl_Panagiotopoulos:03}.


The dependence of the reduced critical density $\rho_\textrm{c}^{*}=\rho_\textrm{c}\sigma^{3}$ on the parameter $\alpha^{-1}$ is
shown in figure~\ref{fig4}.
For very small $\alpha^{-1}$, $\rho_\textrm{c}^{*}$ increases slowly from the RPM value $\rho_\textrm{c}^{*}\simeq 0.009$ and
then at $\alpha^{-1}\simeq 0.048$ changes sharply to the value about $20$ times larger than the RPM limit. For $\alpha^{-1}> 0.048$,
the critical density approaches the SW limit $\rho_\textrm{c}^{*}=0.2457$ passing through a shallow minimum located at $\alpha^{-1}\simeq 0.1$.
Just as standard mean-field theories, the one-loop approximation considered in this paper underestimates the critical number density,
the deviation from the simulation data increases when the strength of the Coulomb interactions becomes sufficiently large \cite{Kristof_Boda:03,Cai-Mol1,Patsahan_Patsahan:10}. At the same time, our results
demonstrate the  behaviour of $\rho_\textrm{c}^{*}$ for small $\alpha^{-1}$   which  qualitatively differs
from the available results of computer simulations (see \cite{Diehl_Panagiotopoulos:03}). In simulations,   the critical density
rather rapidly increases with
$\alpha^{-1}$ in the region $\alpha^{-1}\leqslant 0.1$
but it does not exhibit a jump.   It is worth noting that the behaviour of  $\rho_\textrm{c}^{*}$  for $\alpha^{-1}> 0.1$ qualitatively
agrees with the results of simulations.

It is instructive to view  the  range of the effective density-density  interactions  $R_{N}=\sqrt{a_{2,2}/\sigma}$
[see equation (\ref{a22_sw+pm})] as  a function of $\alpha^{-1}$.
Figure~\ref{fig5} shows the behaviour of $R_{N,\textrm{c}}=R_{N}(T_\textrm{c},\rho_\textrm{c})$  with the variation of $\alpha^{-1}$. The dimensionless
amplitude of the density correlation length
\begin{displaymath}
\xi_{0}^{*}=\frac{\xi_{0}}{\sigma}=\frac{1}{\sigma}\sqrt{\frac{a_{2,2}}{a_{2,t}}}\,,
\qquad
a_{2,t}=\left.\frac{\partial
a_{2,0}}{\partial t}\right|_{t=0}
\end{displaymath}
as a function of $\alpha^{-1}$ is shown in the inset.  Remarkably,
the trends of $R_{N,\textrm{c}}$ and $\rho_\textrm{c}^{*}$ are
very similar, especially for $\alpha^{-1}> 0.048$. Furthermore, the behaviour of  $R_{N,\textrm{c}}$
clearly shows
the two distinctive regions  separated by $\alpha^{-1}\simeq 0.048$:
the region where $R_{N,\textrm{c}}\lesssim 0.5$ and the region where  $R_{N,\textrm{c}}> 0.6$.
Similarly, the two distinctive branches  are clearly seen in the behaviour of  $\xi_{0}^{*}$.
This implies that  $\alpha^{-1}\simeq 0.048$ ($\alpha\simeq 21$) separates the models which demonstrate non-Coulombic
phase behaviour (``solvophobic systems'') from the models   demonstrating Coulombic phase behaviour (``Coulombic systems'').


 \begin{figure}[!t]
 \centering
 \includegraphics[width=0.5\textwidth]{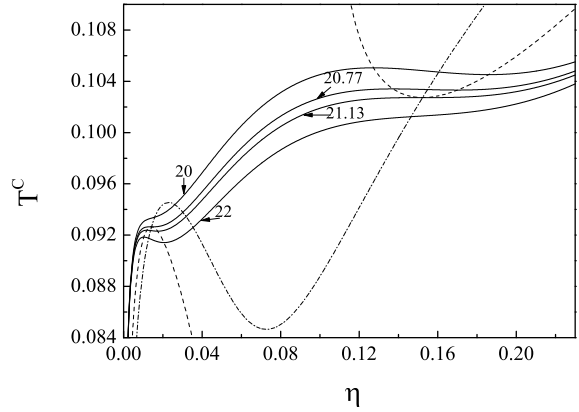}
 \caption{RPM-SW models with  $\alpha=20$, $\alpha=20.77$, $\alpha=21.13$, and $\alpha=22$: the loci of equations $a_{2,0}=0$
 (solid lines),  $a_{3,0}=0$ (dashed lines) and $a_{4,0}=0$ (dash-dotted line).
 $T^\textrm{C}$ is given by equation (\ref{T-C}),
 $\eta=\pi\rho\sigma^{3}/6$ is the packing fraction and $\alpha=q_{0}^{2}z/(\epsilon\sigma\varepsilon)$.
 } \label{fig6}
 \end{figure}

\begin{table}[!b]
\caption{The reduced   parameters of the tricritical points for the  RPM-SW model. The reduced critical temperature $T^{\text{C}}$
is given by (\ref{T-C}) and $T^{\text{SR}}=k_\textrm{B}T/\varepsilon$.\label{tab}}
\vspace{2ex}
\begin{center}
\begin{tabular}{cccc}
\hline \hline\hspace{5mm}  $\alpha$\hspace{5mm} &\hspace{1mm}
$T_{\textrm{trc}}^{\text{C}}$\hspace{6mm}&\hspace{6mm}
$T_{\textrm{trc}}^{\text{SR}}$\hspace{6mm} &\hspace{5mm}
$\rho_{\textrm{trc}}^{*}$\hspace{8mm}
\\
\hline
20.77  & $0.0926$ &$1.9233$ & $0.0269$ \\
21.13 & $0.1027$&$2.1701$ &$0.2905$  \\
\hline \hline
\end{tabular}
\end{center}
\end{table}

  Let us scrupulously consider the RPM-SW models with $\alpha$ close to $\alpha=\alpha^{*}\simeq 21$ ($(\alpha^{*})^{-1}\simeq 0.048$). To this end,  we solve  the equation
\begin{equation}
a_{4,0}(\rho_\textrm{c},T_\textrm{c})  =0
\label{eq_tr}
\end{equation}
additionally to the system of equations (\ref{cr-point}).
The results  are shown in figure~\ref{fig6}.   The solid curves in the figure are the spinodals of the RPM-SW models with
$\alpha=20$, $20.77$, $21.13$, and $22$.
The two dashed curves, a curve with a maximum in the  region of low temperatures  and low densities and another curve with a
minimum in the region of  higher temperatures and moderate densities, indicate the loci of equation $a_{3,0}=0$.
The spinodal of the RPM-SW model with $\alpha=22$  intersects the low-density dashed curve while the spinodal
corresponding to $\alpha=20$ intersects the moderate-density dashed curve. For $\alpha=21.13$, the spinodal   intersects the low-density
curve and it is tangent to the curve placed at higher densities.   The opposite situation appears for $\alpha=20.77$:
the spinodal is tangent to the low-density dashed curve  and intersects another dashed curve located at higher densities.
 The dash-dotted curve which presents the loci of equation (\ref{eq_tr}) crosses  the dashed curves at their extremum points
 (exactly at the points of tangency of the solid and dashed curves). These special points, at which   equations (\ref{cr-point})
 and equation~(\ref{eq_tr}) hold are  called tricritical points \cite{Goldenfeld}. The coordinates of the tricritical points are
 presented in table~\ref{tab}.  The two tricritical points  can be considered as limiting points   which separate the  low-density
 family of gas--liquid phase diagrams from  the gas--liquid phase diagrams located
at moderate densities.    It should be noted that  both a  critical point  and a tricritical point were predicted theoretically
\cite{ciach_stell:2001,Ciach_Stell:2002,ciach:05} and found in computer simulations \cite{Diehl_Panagiotopoulos:03} for the lattice RPM
supplemented by short-range interactions. In these works, the tricritical point is associated with the transition to the
charge-ordered phase. Besides, the tricritical point studied in  \cite{ciach_stell:2001,Ciach_Stell:2002,ciach:05} is either
stable or metastable. On the other hand, no tricritical point was found  in  \cite{Diehl_Panagiotopoulos:03} for the lattice version
of the RPM-SW model with the lattice discretization parameter $\zeta=10$.
According to \cite{Panagiotopoulos_Kumar:99}, for $\zeta\geqslant 3$ the phase behaviour of the lattice RPM is qualitatively identical
to the continuous model. Here, we focus our attention only on the gas--liquid equilibrium. The description of a possible existence of
other phases requires going beyond the treatment  presented  in this work.

 \begin{figure}[!t]
 \centerline{
 \includegraphics[width=0.49\textwidth]{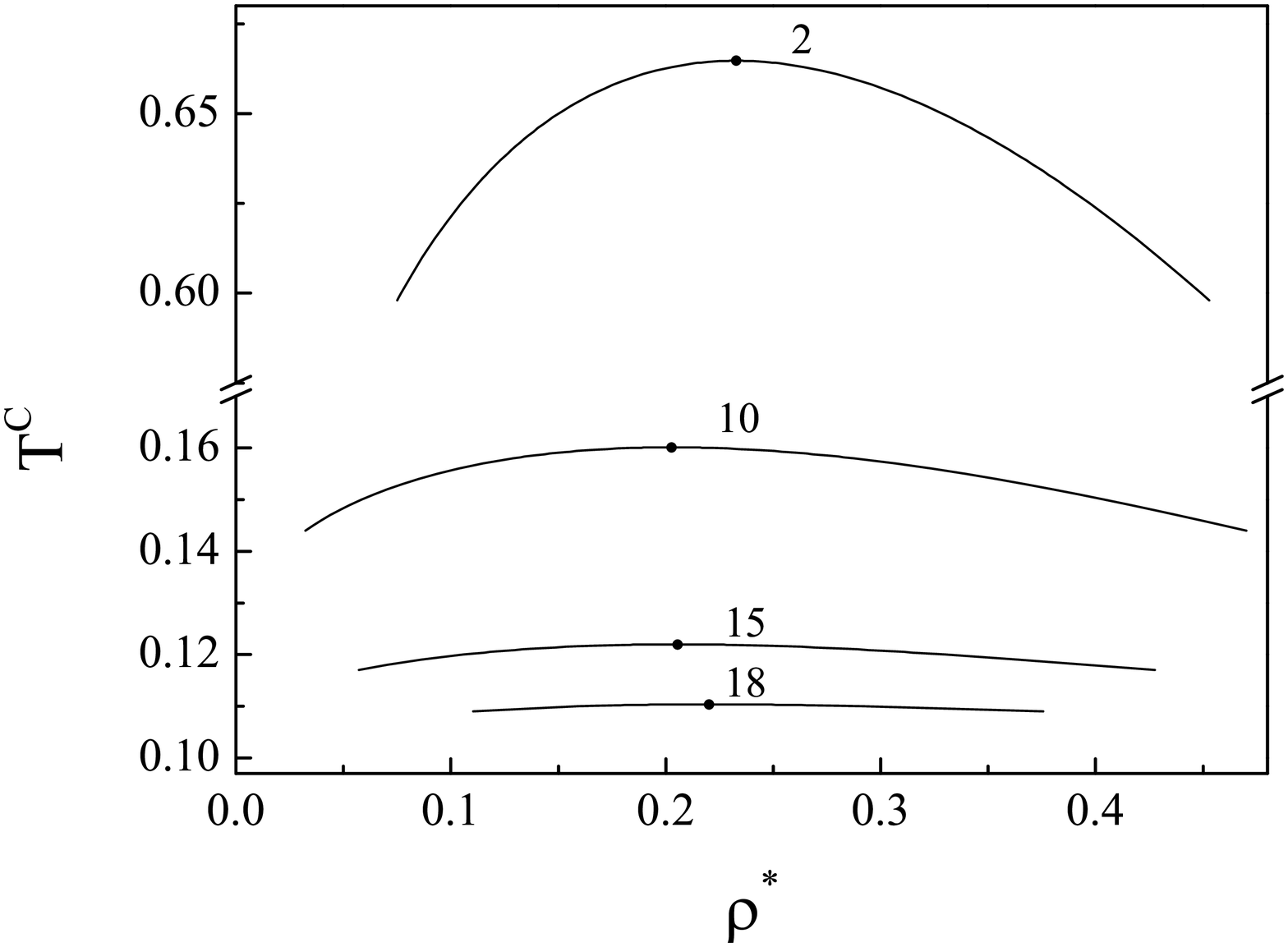}
 \hfill
 \includegraphics[width=0.5\textwidth]{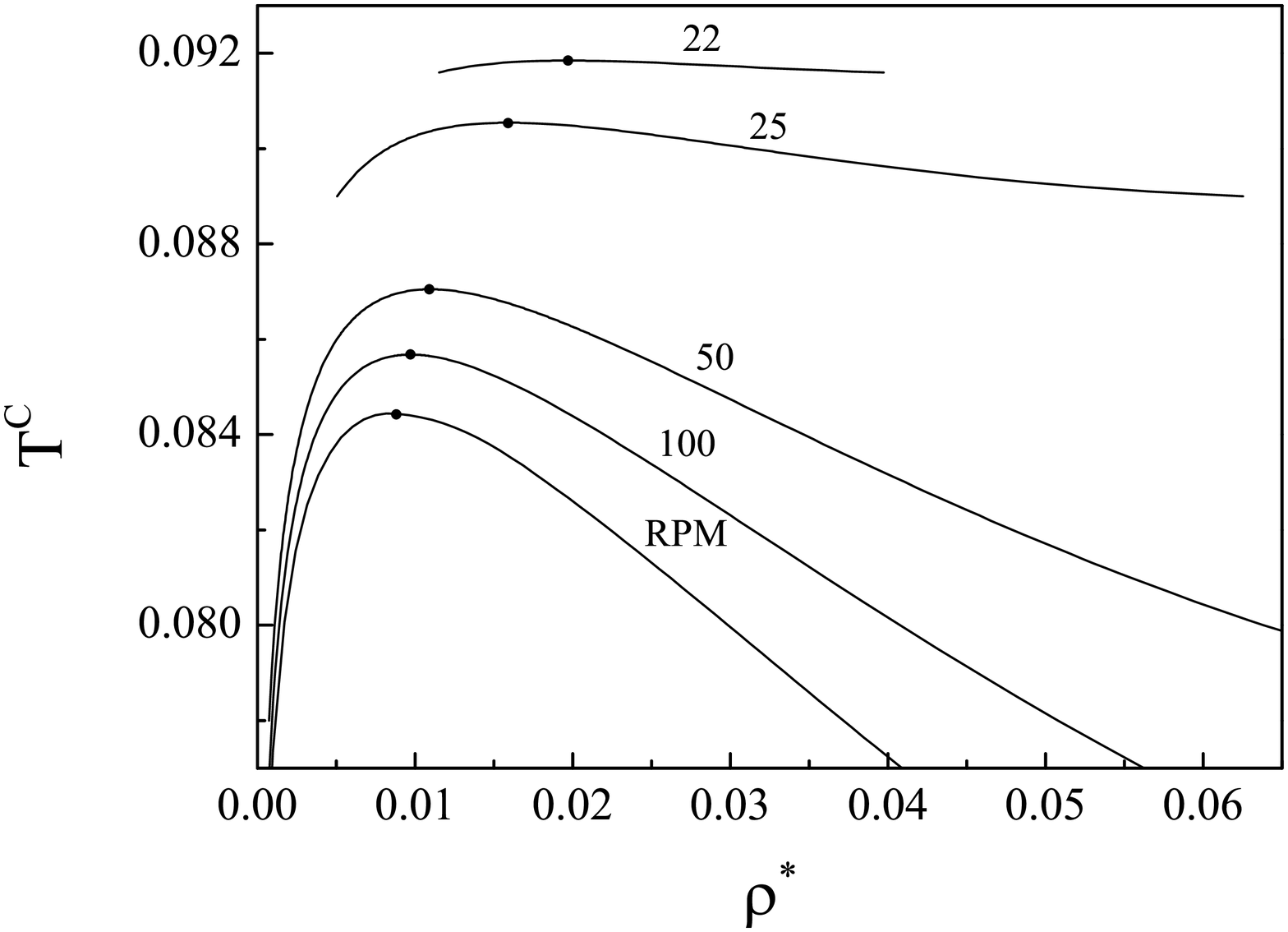}
 }
 \parbox[t]{0.5\textwidth}{
 \caption{Gas--liquid coexistence  curves  of the RPM-SW model for  $\alpha=2$, $10$ and $15$
  in the ($T^\textrm{C}$, $\rho^{*}$)
 representation. Full circles denote the location of the critical point.
 } \label{fig7}
 }
\parbox[t]{0.5\textwidth}{
 \caption{Gas--liquid coexistence  curves of the RPM-SW model for  $\alpha=22$, $25$, $50$ and $100$
  in the ($T^\textrm{C}$, $\rho^{*}$)
 representation. Full circles denote the location of the critical point.
 } \label{fig8}
}
 \end{figure}


Now we  study the evolution of the gas--liquid coexistence curves with the variation of $\alpha$. To this end,  we use  equation~(\ref{a1_sw+pm}) for the  calculation of the chemical potential and  employ the Maxwell
 double-tangent construction. Figures~\ref{fig7} and \ref{fig8} show  the  coexistence curves of the RPM-SW models   in the ($T^\textrm{C}$, $\rho^{*}$) plane
 for two sets of  $\alpha$ corresponding to $\alpha<\alpha^{*}$ and $\alpha>\alpha^{*}$, respectively.
 Figure~\ref{fig7} shows almost symmetrical coexistence envelopes typical of non-Coulombic systems. In  figure~\ref{fig8}, the coexistence curves
 become more and more asymmetric  with an increase of $\alpha$ and resemble  the RPM coexistence curve at $\alpha>100$.
 On the other hand, the coexistence region of the RPM-SW model
 reduces when $\alpha$ tends to  $\alpha^{*}$  (from the both sides) indicating the approach of the tricritical point.

 Therefore, our results suggest that the RPM-SW models with $\alpha\leqslant 20.77$  belong to the non-Coulombic systems in which the phase separation is driven by short-range forces while the RPM-SW
 models with $\alpha\geqslant 21.13$  belong to the Coulomb dominated systems.
 The two families of models  are separated  by a very narrow region of $\alpha$ bounded by   the  tricritical points.

\section{Conclusions}

We have studied the effect of the strength of the Coulomb
interaction on the gas--liquid phase diagram of a  model fluid. The model, referred to as  RPM-SW model, consists of oppositely charged hard spheres of the same diameter with additional short-range attractive interactions.
The short-range attraction is chosen  in the form of the SW potential  of the range $\lambda=1.5\sigma$.
Having introduced the parameter $\alpha$ that determines the strength of the Coulomb interaction with respect to the short-range SW interaction,
we  calculate the  gas--liquid phase diagrams for   $\alpha$ varying  from  $0$ (the purely nonionic system) to $\infty$ (the purely Culombic system). It is worth noting that the parameter $\alpha$ is proportional to the inverse
dielectric constant  of the solvent $\epsilon$  when the ion charges are fixed.

Both the coexistence envelopes and the critical parameters have been calculated in a one-loop approximation which is equivalent to the random phase approximation. We have found  that the very narrow region of $\alpha$ ($20.77\leqslant\alpha\leqslant 21.13$) separates the models demonstrating a nonionic type of  phase behaviour from the models which demonstrate  a ``Coulombic'' type of phase diagrams. This region is  bounded from the both sides by the   tricritical points: one point located at $T_{\textrm{trc},1}^{\text{SR}}=1.9233$ and $\rho_{\textrm{trc},1}^{*}=0.0269$ and another point with the coordinates  $T_{\textrm{trc},2}^{\text{SR}}=2.1701$ and $\rho_{\textrm{trc},2}^{*}=0.2905$.
The dependence of the reduced critical temperature  on $\alpha$ ($\alpha^{-1}$) indicates a continuous variation from a phase transition driven by Coulomb interactions to a transition determined by ``solvophobic'' interactions.
 Such behaviour agrees with the available data of computer simulations performed for the lattice version of the RPM-SW model with high values of the lattice discretization parameter $\zeta$ ($\zeta=\sigma/l=10$) \cite{Diehl_Panagiotopoulos:03}.
On the other hand, the reduced critical density $\rho_\textrm{c}^{*}$   sharply changes with $\alpha$ in the  region between the two tricritical points. This is at variance with  the  results of computer simulations mentioned above. Furthermore,   no tricritical points were found   for the lattice version of the model in the high $\zeta$ case in the computer simulations

We have  studied   the range of the effective density-density interaction $R_{N,\textrm{c}}$ with the variation of $\alpha$ ($\alpha^{-1}$). The results clearly indicate the two distinctive regions in the phase diagram ($R_{N,\textrm{c}}\lesssim 0.5$ and $R_{N,\textrm{c}}\gtrsim 0.6$) which are  separated by $\alpha\simeq 21$. The amplitude of the density correlation length $\xi_{0}^{*}$ also shows  different  trends for $\alpha<21$ and $\alpha>21$. It should be noted that the values of $\xi_{0}^{*}$ are larger for the Coulomb-dominated systems  than for the non-Coulombic systems.    The opposite situation takes place for $R_{N,\textrm{c}}$.

Thus, we have obtained the results for a  phase behaviour of the RPM-SW model that  qualitatively differ from the findings of
the available computer simulations. Further investigations are needed in order to confirm (or rule out) the presence  of the
tricritical points in the  model with $\alpha\simeq 21$.
In particular,  higher-order  approximations for the coefficients of the effective Hamiltonian (\ref{H_eff}) should be considered.
 On the other hand, it would be useful to  perform  computer simulations for the model  parameters  for which  the theory predicts  tricritical points.

\ukrainianpart

\title{Фазова поведінка газ--рідина в іонних плинах: кулонівські проти некулонівських взаємодій}
\author{О. Пацаган}
\address{Інститут фізики конденсованих систем НАН України, вул. І. Свєнціцького, 1, 79011 Львів, Україна}

\makeukrtitle

\begin{abstract}
Використовуючи теорію колективних змінних, вивчається  вплив конкуренції кулонівських і дисперсних сил на фазову
поведінку газ--рідина модельного іонного плину, а саме, зарядо\-асиметричної примітивної моделі з додатковими
короткосяжними притягальними взаємодіями.  Отримано в однопетлевому наближенні критичні параметри і криві співіснування
в залежності від параметра $\alpha$, який вимірює  силу кулонівської взаємодії
по відношенню до короткосяжної взаємодії. Знайдено дуже вузьку область $\alpha$, обмежену з обох сторін трикритичними точками, яка
відокремлює моделі з ``неіонною'' і ``кулонівською'' фазовою поведінкою. Цей результат  відрізняється від наявного  результату
комп'ютерного моделювання, отриманого для дрібнодискретизованої граткової версії моделі, що розглядається.
\keywords іонні плини, фазова діаграма газ--рідина, трикритична точка, кулонівські взаємодії, короткосяжне притягання

\end{abstract}

\end{document}